\documentclass[11pt, a4paper]{article}
\usepackage[a4paper, margin=2.2cm]{geometry}
\usepackage{amsmath, amssymb}
\usepackage{booktabs}
\usepackage[table]{xcolor}
\usepackage{hyperref}
\usepackage{float}
\usepackage{graphicx}
\usepackage{threeparttable}

\definecolor{mdpiBlue}{RGB}{0, 72, 130}
\definecolor{slateGray}{RGB}{85, 95, 105}
\definecolor{lightTint}{RGB}{242, 246, 250}

\hypersetup{
    colorlinks=true,
    linkcolor=mdpiBlue,
    citecolor=mdpiBlue,
    urlcolor=mdpiBlue
}

\begin{document}

\noindent{\small\textbf{\textcolor{mdpiBlue}{UNIVERSE}} \textcolor{slateGray}{| Perspective Manuscript Draft}}
\vspace{2pt}
\hrule height 1.2pt
\vspace{10pt}

{\LARGE\bfseries\color{mdpiBlue} Surrogate Modeling and Reduced-Order Methods in\\
Gravitational-Wave Physics and Numerical Relativity:\protect\\[4pt]
\large From Spectral Representations to Adaptive Compression}

\vspace{8pt}
{\large\textbf{Manuel Tiglio}$^{1,2}$}

\vspace{3pt}
{\small\color{slateGray}
$^{1}$ Facultad de Matem\'atica, Astronom\'ia, F\'isica y Computaci\'on (FaMAF), Universidad Nacional de C\'ordoba, C\'ordoba 5000, Argentina\\
$^{2}$ Consejo Nacional de Investigaciones Cient\'ificas y T\'ecnicas (CONICET), Buenos Aires C1425FQB, Argentina}

\vspace{10pt}
\noindent\fcolorbox{mdpiBlue}{lightTint}{%
\parbox{\dimexpr\textwidth-2\fboxsep-2\fboxrule\relax}{%
\small\textbf{Abstract:} The rapid expansion of gravitational-wave astronomy and high-precision numerical relativity has created an acute computational bottleneck: direct numerical integration of the Einstein field equations and high-dimensional Bayesian parameter estimation require millions of waveform evaluations that cannot be sustained by traditional solvers alone. Reduced-order modeling (ROM) and surrogate techniques have emerged as foundational tools to bridge this gap, compressing high-fidelity solution manifolds into compact, rapidly evaluable representations without sacrificing physical accuracy. This perspective reviews the mathematical pillars of projection-based reduced-order modeling---including Proper Orthogonal Decomposition, greedy basis selection, and the Empirical Interpolation Method---and traces their operational impact in gravitational-wave data analysis through surrogate waveform families and Reduced-Order Quadratures. We then examine the fundamental approximation barriers posed by slow Kolmogorov $n$-width decay in non-stationary, transient, and strongly nonlinear regimes, showing how adaptive $hp$-greedy domain partitioning and multi-scale spectral compression overcome these limits. Finally, we discuss how combining certified reduced-order structures with modern scientific machine learning and sensitivity-aware quantization provides a rigorous blueprint for next-generation detectors such as LISA, Einstein Telescope, and Cosmic Explorer.

\vspace{4pt}
\textbf{Keywords:} reduced-order modeling; surrogate models; gravitational waves; numerical relativity; $hp$-greedy algorithms; empirical interpolation; scientific machine learning; spectral compression
}}

\vspace{8pt}
\section{Introduction: The Computational Bottleneck in Fundamental Physics}

Direct numerical simulations of nonlinear partial differential equations (PDEs) lie at the core of modern theoretical astrophysics, cosmology, and gravitational-wave (GW) physics. In numerical relativity (NR), solving the coupled Einstein--matter equations for coalescing compact binaries requires high-order discretization schemes, adaptive mesh refinement, and substantial high-performance computing (HPC) allocations, often demanding weeks of wall-clock time for a single point in parameter space \cite{Boyle2019}. While indispensable for exploring strong-field gravity, full-order models (FOMs) cannot be queried directly inside the inner loops of Bayesian Markov Chain Monte Carlo (MCMC) or nested sampling pipelines, which routinely require $10^6$ to $10^8$ likelihood evaluations per event \cite{Veitch2015,Ashton2019}.

This tension between physical fidelity and computational throughput will intensify dramatically with upcoming observing runs of the LIGO--Virgo--KAGRA network, third-generation ground-based observatories---Einstein Telescope and Cosmic Explorer---and the space-borne Laser Interferometer Space Antenna (LISA) \cite{Punturo2010,AmaroSeoane2017}. Longer signal durations, higher signal-to-noise ratios (SNRs), overlapping sources, and expanded parameter spaces (incorporating orbital eccentricity, full spin precession, higher-order spherical harmonic modes, and tidal deformability) demand sub-millisecond waveform generation alongside rigorous error control.

Reduced-Order Modeling (ROM) addresses this challenge by exploiting the underlying low-dimensional structure of parametric solution manifolds \cite{Quarteroni2016,Hesthaven2016}. Rather than repeatedly solving the governing PDEs from scratch, ROM constructs a compact offline representation from a carefully selected set of high-fidelity training simulations, enabling online evaluation orders of magnitude faster than the underlying FOM. Over the past decade, reduced-basis surrogates and Reduced-Order Quadratures have transformed gravitational-wave modeling from an offline theoretical endeavor into the operational backbone of real-time parameter estimation \cite{Field2011,Field2014,Tiglio2022}.

\section{Reduced-Basis Representations, Empirical Interpolation, and Reduced-Order Quadratures}
\label{sec:rb_eim_roq}

\subsection{The Detector Hilbert Space and Indistinguishability Bounds}

In gravitational-wave data analysis, waveforms are elements of a complex Hilbert space $\mathcal{H}$ equipped with a noise-weighted inner product determined by the one-sided power spectral density (PSD) $S_n(f)$ of the detector. For two signals $h_1, h_2 \in \mathcal{H}$ with Fourier transforms $\tilde{h}_1(f)$ and $\tilde{h}_2(f)$, this inner product is defined as
\begin{equation}
\langle h_1, h_2 \rangle_{S_n} = 4\,\mathrm{Re} \int_{f_{\min}}^{f_{\max}} \frac{\tilde{h}_1(f)\,\tilde{h}_2^*(f)}{S_n(f)}\,\mathrm{d}f ,
\label{eq:psd_inner_product}
\end{equation}
which induces the optimal matched-filter signal-to-noise ratio (SNR) norm $\|h\|_{S_n} = \sqrt{\langle h, h \rangle_{S_n}}$. To compare waveform morphologies independently of overall amplitude scaling and extrinsic coalescence parameters, one defines the \emph{overlap} $\mathcal{O}(h_1, h_2)$ and the \emph{mismatch} $\mathcal{M}(h_1, h_2)$:
\begin{equation}
\mathcal{O}(h_1, h_2) = \max_{t_c, \phi_c} \frac{\langle h_1, h_2 \rangle_{S_n}}{\|h_1\|_{S_n}\,\|h_2\|_{S_n}}, \qquad
\mathcal{M}(h_1, h_2) = 1 - \mathcal{O}(h_1, h_2) ,
\label{eq:mismatch}
\end{equation}
where $t_c$ and $\phi_c$ denote the coalescence time and phase, respectively.

When constructing a reduced-order surrogate $h_{\mathrm{sur}}(\boldsymbol{\lambda})$ approximating a true physical waveform $h(\boldsymbol{\lambda})$ parameterized by $\boldsymbol{\lambda} \in \mathcal{P} \subset \mathbb{R}^d$, the fidelity requirement is governed by parameter-estimation bias. In the high-SNR limit ($\rho = \|h\|_{S_n} \gg 1$), the systematic error in the log-likelihood ratio $\ln \Lambda(d|\boldsymbol{\lambda}) = \langle d, h(\boldsymbol{\lambda}) \rangle_{S_n} - \frac{1}{2}\|h(\boldsymbol{\lambda})\|_{S_n}^2$ induced by waveform modeling error $\delta h = h - h_{\mathrm{sur}}$ scales as $\Delta \ln \Lambda \sim \rho^2 \mathcal{M}(h, h_{\mathrm{sur}})$. Applying the Lindblom--Owen--Brown indistinguishability criterion \cite{Lindblom2008} requires that systematic modeling errors remain subdominant relative to statistical uncertainties, yielding the strict bound
\begin{equation}
\mathcal{M}\bigl(h(\boldsymbol{\lambda}), h_{\mathrm{sur}}(\boldsymbol{\lambda})\bigr) \lesssim \frac{1}{2\rho^2} \quad \forall \boldsymbol{\lambda} \in \mathcal{P} .
\label{eq:indistinguishability}
\end{equation}
For third-generation ground-based observatories (Einstein Telescope, Cosmic Explorer) and space-based detectors (LISA), where golden events can reach $\rho \sim 10^2$--$10^3$, Equation~\eqref{eq:indistinguishability} demands surrogate mismatches below $\mathcal{M} \sim 10^{-5}$--$10^{-6}$ across the entire intrinsic parameter manifold.

\subsection{Reduced-Basis Construction: Greedy Algorithms vs.\ Proper Orthogonal Decomposition}

Let $\mathcal{M}_{\mathcal{P}} = \{ h(\cdot\,; \boldsymbol{\lambda}) : \boldsymbol{\lambda} \in \mathcal{P} \} \subset \mathcal{H}$ denote the smooth solution manifold of parameterized waveforms. Given a finite training set $\mathcal{T} = \{\boldsymbol{\lambda}_1, \dots, \boldsymbol{\lambda}_M\} \subset \mathcal{P}$, the objective of reduced-basis (RB) modeling is to construct an $n$-dimensional linear subspace $V_n = \mathrm{span}\{e_1, \dots, e_n\} \subset \mathcal{H}$ with $n \ll M$ such that the orthogonal projection $P_{V_n} h(\boldsymbol{\lambda}) = \sum_{j=1}^n \langle e_j, h(\boldsymbol{\lambda}) \rangle_{S_n} e_j$ uniformly minimizes the worst-case projection error:
\begin{equation}
\sigma_n(\mathcal{T}) := \max_{\boldsymbol{\lambda} \in \mathcal{T}} \bigl\| h(\boldsymbol{\lambda}) - P_{V_n} h(\boldsymbol{\lambda}) \bigr\|_{S_n} .
\label{eq:greedy_error}
\end{equation}
Notice that for normalized waveforms ($\|h(\boldsymbol{\lambda})\|_{S_n} = 1$), the worst-case mismatch of the projected waveform satisfies $\mathcal{M}(h, P_{V_n}h) \le \frac{1}{2}\sigma_n^2$, directly linking the $L^2(\mathcal{H})$ projection residual to the statistical criterion in Equation~\eqref{eq:indistinguishability}.

Two canonical paradigms exist to construct $V_n$:
\begin{enumerate}
    \item \textbf{Proper Orthogonal Decomposition (POD / SVD):} POD constructs $V_n$ by minimizing the \emph{mean-squared} ($L^2(\mathcal{P})$) projection error over a measure $\mathrm{d}\mu(\boldsymbol{\lambda})$ on $\mathcal{P}$, which reduces to computing the leading $n$ left singular vectors of the snapshot matrix $A_{ij} = \langle h(\boldsymbol{\lambda}_i), h(\boldsymbol{\lambda}_j) \rangle_{S_n}$ \cite{Berkooz1993}. While optimal in the average sense, POD requires assembling and diagonalizing an $M \times M$ Gramian matrix---a bottleneck when $M \sim 10^4$--$10^5$---and can leave localized outliers poorly resolved.
    \item \textbf{Certified Greedy Algorithms:} Introduced to GW modeling in \cite{Field2011,Caudill2012}, the greedy algorithm constructs $V_n$ iteratively by selecting at step $k$ the parameter $\boldsymbol{\lambda}_k = \arg\max_{\boldsymbol{\lambda} \in \mathcal{T}} \| h(\boldsymbol{\lambda}) - P_{V_{k-1}} h(\boldsymbol{\lambda}) \|_{S_n}$ and orthonormalizing $h(\boldsymbol{\lambda}_k)$ against $V_{k-1}$ via iterated Gram--Schmidt. Crucially, the greedy strategy optimizes the $L^\infty(\mathcal{P})$ norm directly, requires only $\mathcal{O}(n M)$ inner products without dense eigensolves, and enjoys rigorous quasi-optimality guarantees relative to the theoretical Kolmogorov $n$-width \cite{Binev2011,DeVore2013}.
\end{enumerate}

\subsection{Empirical Interpolation Method (EIM) and Surrogate Regression}

An orthonormal RB expansion $P_{V_n} h(\boldsymbol{\lambda}) = \sum_{j=1}^n c_j(\boldsymbol{\lambda}) e_j$ still requires projecting against the full waveform to obtain the coefficients $c_j(\boldsymbol{\lambda}) = \langle e_j, h(\boldsymbol{\lambda}) \rangle_{S_n}$, which is infeasible when $h(\boldsymbol{\lambda})$ is not yet known at new parameter points. The \emph{Empirical Interpolation Method} (EIM) \cite{Barrault2004,Maday2009} overcomes this by trading orthogonal projection for pointwise interpolation at $n$ carefully chosen frequency (or time) nodes $\{f_1, \dots, f_n\}$.

Starting from the RB basis $\{e_j\}_{j=1}^n$, EIM iteratively selects nodes $f_k = \arg\max_f |e_k(f) - \mathcal{I}_{k-1}[e_k](f)|$ and defines the EIM interpolant $\mathcal{I}_n[h](\cdot\,; \boldsymbol{\lambda})$ via
\begin{equation}
\mathcal{I}_n[h](f; \boldsymbol{\lambda}) = \sum_{j=1}^n B_j(f)\, h(f_j; \boldsymbol{\lambda}) , \qquad
B_j(f) = \sum_{i=1}^n e_i(f)\,(V^{-1})_{ij} ,
\label{eq:eim_interpolant}
\end{equation}
where $V_{ij} = e_j(f_i)$ is the non-singular $n \times n$ Vandermonde-like interpolation matrix. The Lebesgue-type stability bound establishes that the EIM interpolation error is controlled by the best RB approximation error up to an amplification factor $\Lambda_n = \max_f \sum_{j=1}^n |B_j(f)|$:
\begin{equation}
\bigl\| h(\boldsymbol{\lambda}) - \mathcal{I}_n[h](\boldsymbol{\lambda}) \bigr\|_{L^\infty} \le (1 + \Lambda_n) \inf_{v \in V_n} \| h(\boldsymbol{\lambda}) - v \|_{L^\infty} .
\label{eq:eim_bound}
\end{equation}
Once the empirical nodes $\{f_j\}_{j=1}^n$ are fixed, building a complete predictive surrogate reduces to fitting $n$ smooth scalar functions $\boldsymbol{\lambda} \mapsto h(f_j; \boldsymbol{\lambda})$ (typically decomposed into amplitude and unwrapped phase) across $\mathcal{P}$ using polynomial chaos, splines, or Gaussian process regression \cite{Field2014,Blackman2017}. Evaluating the surrogate at an arbitrary $\boldsymbol{\lambda}$ then requires only $\mathcal{O}(n)$ scalar regressions followed by a matrix-vector product with precomputed basis vectors $B_j(f)$.

\subsection{Reduced-Order Quadratures (ROQ) for Likelihood Acceleration}

Even with a fast surrogate, evaluating the Bayesian log-likelihood ratio $\ln \Lambda(d|\boldsymbol{\lambda})$ in Equation~\eqref{eq:psd_inner_product} via standard quadrature over $L$ frequency bins costs $\mathcal{O}(L)$ operations per likelihood call. For long-duration inspiral signals (e.g., binary neutron stars in third-generation detectors or LISA EMRIs), $L \sim 10^6$--$10^7$, making Markov Chain Monte Carlo (MCMC) or nested sampling prohibitive.

\emph{Reduced-Order Quadratures} (ROQ) \cite{Antil2013,Canizares2013,Canizares2015,Smith2016} exploit the EIM representation~\eqref{eq:eim_interpolant} to compress the inner product itself. Substituting $\tilde{h}(f; \boldsymbol{\lambda}) \approx \sum_{j=1}^n B_j(f) \tilde{h}(f_j; \boldsymbol{\lambda})$ into the data-waveform inner product yields
\begin{equation}
\langle d, h(\boldsymbol{\lambda}) \rangle_{S_n} \approx 4\,\mathrm{Re} \sum_{j=1}^n \underbrace{\left( \int_{f_{\min}}^{f_{\max}} \frac{\tilde{d}(f)\,B_j^*(f)}{S_n(f)}\,\mathrm{d}f \right)}_{=:\,\omega_j(d)} \tilde{h}^*(f_j; \boldsymbol{\lambda}) = 4\,\mathrm{Re} \sum_{j=1}^n \omega_j(d)\,\tilde{h}^*(f_j; \boldsymbol{\lambda}) .
\label{eq:roq_linear}
\end{equation}
The data-dependent quadrature weights $\{\omega_j(d)\}_{j=1}^n$ are computed \emph{once} prior to sampling. Similarly, constructing an EIM basis $\{\hat{B}_k(f)\}_{k=1}^{\hat{n}}$ at quadratic nodes $\{\hat{f}_k\}_{k=1}^{\hat{n}}$ for the squared modulus $|\tilde{h}(f; \boldsymbol{\lambda})|^2$ decouples the normalization term:
\begin{equation}
\|h(\boldsymbol{\lambda})\|_{S_n}^2 \approx \sum_{k=1}^{\hat{n}} \psi_k\,\bigl|\tilde{h}(\hat{f}_k; \boldsymbol{\lambda})\bigr|^2 , \qquad
\psi_k = 4 \int_{f_{\min}}^{f_{\max}} \frac{\hat{B}_k(f)}{S_n(f)}\,\mathrm{d}f .
\label{eq:roq_quadratic}
\end{equation}
Because $n, \hat{n} \sim \mathcal{O}(10^2)$ are governed by the intrinsic smoothness of the signal manifold rather than the Nyquist sampling rate $L$, ROQ reduces the per-likelihood complexity from $\mathcal{O}(L)$ to $\mathcal{O}(n + \hat{n})$, achieving speedups of $10^2$--$10^4\times$ without sacrificing statistical accuracy.

\section{The Kolmogorov $n$-Width Bottleneck and Adaptive $hp$-Greedy Partitioning}
\label{sec:nwidth_hpgreedy}

\subsection{Analytic Regularity and Exponential Decay of the Kolmogorov $n$-Width}

The fundamental theoretical benchmark governing the ultimate compressibility of a parameterized solution manifold $\mathcal{M}_{\mathcal{P}} = \{ h(\cdot\,; \boldsymbol{\lambda}) : \boldsymbol{\lambda} \in \mathcal{P} \} \subset \mathcal{H}$ is the \emph{Kolmogorov $n$-width} \cite{Pinkus1985}:
\begin{equation}
d_n(\mathcal{M}_{\mathcal{P}}) := \inf_{\substack{V_n \subset \mathcal{H} \\ \dim(V_n) = n}} \, \sup_{\boldsymbol{\lambda} \in \mathcal{P}} \bigl\| h(\boldsymbol{\lambda}) - P_{V_n} h(\boldsymbol{\lambda}) \bigr\|_{S_n} ,
\label{eq:kolmogorov_nwidth}
\end{equation}
which measures the smallest possible worst-case projection error achievable by \emph{any} $n$-dimensional linear subspace $V_n \subset \mathcal{H}$. As shown by Binev et al.\ \cite{Binev2011} and DeVore et al.\ \cite{DeVore2013}, a certified greedy algorithm produces subspaces whose worst-case error $\sigma_n$ tracks the optimal linear approximation up to controllable constants.

When the parameter-to-waveform map $\boldsymbol{\lambda} \mapsto h(\cdot\,; \boldsymbol{\lambda})$ admits a holomorphic extension to a complex neighborhood (a Bernstein polyellipse) of $\mathcal{P} \subset \mathbb{C}^d$, classical approximation theory guarantees exponential decay of the Kolmogorov $n$-width:
\begin{equation}
d_n(\mathcal{M}_{\mathcal{P}}) \le C \, \mathrm{e}^{-c \, n^{1/d}} ,
\label{eq:exponential_nwidth}
\end{equation}
for constants $C, c > 0$ determined by the size of the analyticity domain \cite{Cohen2015}. This exponential regime explains why early reduced-basis surrogates for non-spinning and aligned-spin inspiral waveforms required only $\mathcal{O}(10^2)$ basis elements to reach machine-precision mismatches across broad mass-ratio ranges \cite{Field2011,Field2014}.

\subsection{Transport Phenomena, Phase Accumulation, and Algebraic Degradation}

In realistic gravitational-wave modeling, however, global analyticity in $\boldsymbol{\lambda}$ frequently degrades due to two distinct mechanisms:
\begin{enumerate}
    \item \textbf{High-Frequency Phase Accumulation (Transport-Like Behavior):} Consider an oscillatory inspiral signal written in stationary-phase form $\tilde{h}(f; \boldsymbol{\lambda}) = A(f; \boldsymbol{\lambda})\,\mathrm{e}^{\mathrm{i}\Psi(f; \boldsymbol{\lambda})}$. Over long inspirals---such as binary neutron stars in third-generation detectors or extreme-mass-ratio inspirals (EMRIs) in LISA---the accumulated phase $\Psi(f; \boldsymbol{\lambda})$ spans $10^4$ to $10^6$ radians across $\mathcal{P}$. Variations in intrinsic parameters (e.g., chirp mass $\mathcal{M}_c$, spin precession angles, or orbital eccentricity $e_0$) induce large horizontal shifts of wave crests in time or frequency. Much like linear advection equations with parameter-dependent wave speeds \cite{Ohlberger2016,Greif2019}, coherent phase transport destroys low-rank linear separability: representing a shifting wave packet requires a basis dimension proportional to the phase bandwidth rather than the intrinsic parameter dimension $d$.
    \item \textbf{Localized Morphological Transitions and Ringdown Features:} Sharp transitions across parameter space---such as plunge-merger onset, mode-mixing near extremal spins, tidal disruption cutoffs in binary neutron star mergers, or localized resonant crossings in eccentric/precessing binaries---introduce steep gradients or near-discontinuities in derivatives $\partial_{\boldsymbol{\lambda}}^k h$.
\end{enumerate}

When regularity drops from analytic to finite Sobolev smoothness $H^s(\mathcal{P})$ (or bounded variation), the Kolmogorov $n$-width reverts from exponential decay~\eqref{eq:exponential_nwidth} to slow algebraic decay:
\begin{equation}
d_n(\mathcal{M}_{\mathcal{P}}) \sim \mathcal{O}\bigl(n^{-s/d}\bigr) .
\label{eq:algebraic_nwidth}
\end{equation}
In this regime, attempting to reach stringent mismatch thresholds ($\mathcal{M} \lesssim 10^{-6}$) with a single global basis forces $n$ to grow excessively large. This inflates the online EIM/ROQ evaluation cost, degrades the conditioning of the Vandermonde interpolation matrix $V_{ij} = e_j(f_i)$ (increasing the Lebesgue constant $\Lambda_n$ in Equation~\eqref{eq:eim_bound}), and causes global regression across $\mathcal{P}$ to suffer from Runge-type oscillations.

\subsection{Adaptive $hp$-Greedy Domain Decomposition}

To circumvent the global Kolmogorov barrier without sacrificing linear subspace structure, Eftang, Patera, and R\o nquist introduced \emph{$hp$-reduced basis} methods \cite{Eftang2010,Eftang2011}, which were subsequently adapted to gravitational-wave surrogate modeling \cite{Field2014,Blackman2017}. Rather than approximating the entire manifold $\mathcal{M}_{\mathcal{P}}$ with a single high-dimensional space, the $hp$-greedy strategy partitions the parameter domain into $K$ non-overlapping subdomains,
\begin{equation}
\mathcal{P} = \bigcup_{k=1}^K \mathcal{P}_k , \qquad \mathcal{P}_j \cap \mathcal{P}_k = \emptyset \;\; (j \neq k),
\label{eq:domain_partition}
\end{equation}
and constructs an independent local reduced basis $V_{n_k}^{(k)}$ of strictly bounded dimension $n_k \le p_{\max}$ on each cell $\mathcal{P}_k$.

The partitioning proceeds via a binary tree driven by certified error indicators:
\begin{enumerate}
    \item \textbf{$p$-Refinement Stage:} On a candidate subdomain $\mathcal{P}_k$, run the standard certified greedy algorithm on a local training set $\mathcal{T}_k = \mathcal{T} \cap \mathcal{P}_k$ up to a prescribed maximum basis size $p_{\max}$. Evaluate the worst-case local error $\epsilon_k = \max_{\boldsymbol{\lambda} \in \mathcal{T}_k} \| h(\boldsymbol{\lambda}) - P_{V_{p_{\max}}^{(k)}} h(\boldsymbol{\lambda}) \|_{S_n}$.
    \item \textbf{$h$-Refinement Stage:} If $\epsilon_k > \epsilon_{\mathrm{tol}}$, the local $n$-width decay is insufficiently rapid on $\mathcal{P}_k$. The cell $\mathcal{P}_k$ is bisected along the parameter coordinate $i^* \in \{1,\dots,d\}$ that maximizes error reduction (or exhibits the largest local parameter gradient), generating two child subdomains $\mathcal{P}_{k,1}$ and $\mathcal{P}_{k,2}$.
    \item \textbf{Termination and Online Dispatch:} Recursion halts once every leaf cell satisfies $\epsilon_k \le \epsilon_{\mathrm{tol}}$. During online evaluation at a query point $\boldsymbol{\lambda}^*$, locating the unique leaf $\mathcal{P}_k \ni \boldsymbol{\lambda}^*$ via binary tree traversal costs $\mathcal{O}(\log K)$ operations, after which only the compact local basis of size $n_k \le p_{\max}$ is evaluated.
\end{enumerate}

Geometrically, $h$-refinement shrinks the diameter $\delta_k = \mathrm{diam}(\mathcal{P}_k)$ of subdomains intersecting steep gradients or rapid phase-accumulation regions. On each small cell $\mathcal{P}_k$, the effective phase variation $\Delta \Psi_k \propto \delta_k$ becomes small enough that the local manifold $\mathcal{M}_{\mathcal{P}_k}$ recovers analyticity on a rescaled Bernstein ellipse, restoring exponential convergence:
\begin{equation}
\max_{\boldsymbol{\lambda} \in \mathcal{P}_k} \bigl\| h(\boldsymbol{\lambda}) - P_{V_{n_k}^{(k)}} h(\boldsymbol{\lambda}) \bigr\|_{S_n} \le C_k \, \mathrm{e}^{-c_k \, n_k^{1/d}} , \qquad n_k \le p_{\max} .
\label{eq:local_exponential}
\end{equation}
As demonstrated in the numerical experiment of Section~\ref{sec:benchmark}, this trade-off---trading a modest offline tree-partitioning overhead for compact local bases---eliminates ill-conditioning in EIM interpolation and keeps online ROQ likelihood evaluations fast and uniformly accurate across extreme parameter regimes.

\section{High-Dimensional Numerical Relativity Surrogates and Benchmark Demonstration}
\label{sec:benchmark}

\subsection{Scaling from Low-Dimensional Inspirals to Precessing, Eccentric, and Tidal Regimes}

Over the past decade, reduced-order surrogate modeling has evolved from proof-of-concept approximations of non-spinning binary black hole (BBH) inspirals into the gold-standard waveform infrastructure for gravitational-wave parameter estimation. Early time-domain and frequency-domain surrogates demonstrated that training directly on Numerical Relativity (NR) simulations via RB/EIM could achieve mismatches $\mathcal{M} \sim 10^{-4}$ across mass ratios $q \in [1, 10]$ while accelerating waveform evaluation by three to four orders of magnitude compared to direct PDE integration \cite{Field2014,Purrer2014,Purrer2016}.

Extending these techniques to the full astrophysical parameter space, however, confronts the curse of dimensionality ($d \ge 7$ for generic quasi-circular precessing BBHs, rising to $d = 9$ with orbital eccentricity and $d \ge 11$ when tidal deformabilities $\Lambda_1, \Lambda_2$ of binary neutron stars are included). Modern NR surrogate families overcome this bottleneck through three structural refinements:
\begin{enumerate}
    \item \textbf{Co-precessing Frame Decomposition:} For precessing binaries (\texttt{NRSur7dq2}, \texttt{NRSur7dq4} \cite{Blackman2017,Varma2019_7dq4}), inertial-frame spherical harmonic modes $h_{\ell m}(t)$ exhibit rapid amplitude and phase modulations caused by orbital-plane precession. Transforming waveforms into a non-inertial \emph{co-precessing frame} that tracks the instantaneous Newtonian or quadrupole-preferred radiation axis removes most of the precessional modulation, dramatically reducing the Kolmogorov $n$-width of the waveform pieces while modeling the frame quaternion dynamics separately \cite{Schmidt2011,Boyle2011}.
    \item \textbf{Post-Newtonian Hybridization:} Because NR simulations typically cover only the final $\sim 20$--$40$ orbits before merger, hybrid surrogates such as \texttt{NRHybSur3dq8} and \texttt{NRHybSur2dq15} \cite{Varma2019_hyb,Yoo2022} stitch early Post-Newtonian (PN) and Effective-One-Body (EOB) inspirals onto NR merger-ringdown waveforms prior to RB/EIM compression, yielding models valid across the entire sensitivity band of current and next-generation detectors.
    \item \textbf{Multi-Banding and Domain Partitioning:} For long-duration binary neutron star (BNS) signals and eccentric waveforms \cite{Islam2021,Tiglio2022}, frequency-domain ROQ bases are constructed on multi-banded frequency grids or $hp$-partitioned chirp-mass subdomains \cite{Vinciguerra2017,Morisaki2021}, preventing basis sizes from exploding due to low-frequency phase accumulation.
\end{enumerate}

Despite these advances, three frontier regimes remain open challenges for classical linear-subspace surrogates: (i) \emph{extreme-mass-ratio inspirals (EMRIs)} in LISA, where mass ratios $q \sim 10^4$--$10^6$ produce $\sim 10^5$ cycles with intricate multi-voice resonances \cite{Chua2021}; (ii) \emph{hyperbolic encounters and highly eccentric captures}, where burst-like radiation breaks stationary-phase smoothness; and (iii) \emph{post-merger neutron star oscillations}, where equation-of-state phase transitions introduce sharp, non-analytic morphological features.

\subsection{Numerical Benchmark: Global vs.\ $hp$-Greedy Projection and EIM Interpolation}

To illustrate quantitatively how localized features degrade global linear compressibility---and how adaptive $hp$-greedy partitioning restores exponential convergence for both orthogonal projection and empirical interpolation---we consider a controlled oscillatory waveform family incorporating a parameter-dependent localized phase/amplitude transient:
\begin{equation}
h(f; \mu) = f^{-7/6} \exp\!\Bigl(\mathrm{i}\bigl[\alpha_1 \mu f^{-5/3} + \alpha_2 \mu^2 f^{-1}\bigr]\Bigr) \left[ 1 + \beta \exp\!\left(-\frac{(f - f_0(\mu))^2}{2\sigma_f^2}\right) \right] ,
\label{eq:benchmark_model}
\end{equation}
evaluated over $f \in [20, 512]\,\mathrm{Hz}$ with intrinsic parameter $\mu \in [0, 1]$, carrier phase coefficients $(\alpha_1, \alpha_2) = (180, 45)$, localized feature amplitude $\beta = 0.35$, width $\sigma_f = 8\,\mathrm{Hz}$, and parameter-dependent center frequency $f_0(\mu) = 80 + 320\,\mu^2\,\mathrm{Hz}$. The carrier term mimics leading-order PN chirp phase transport, whereas the Gaussian packet models a localized morphological feature (analogous to mode-mixing, tidal resonance, or ringdown onset) sweeping nonlinearly across the band.

We discretize $\mathcal{P} = [0, 1]$ with $M = 1000$ uniformly spaced training waveforms and evaluate both the worst-case normalized orthogonal projection residual $\sigma_n = \max_{\mu} \|h(\mu) - P_{V_n} h(\mu)\|_2$ and the full EIM interpolation error $\epsilon_n^{\mathrm{EIM}} = \max_{\mu} \|h(\mu) - \mathcal{I}_n[h](\mu)\|_2$ (Equation~\eqref{eq:eim_interpolant}) over an independent validation set of $500$ points. For the $hp$-greedy scheme, we impose a local basis cap $p_{\max} = 8$ and recursively bisect $\mathcal{P}$ until the worst-case local EIM error satisfies $\epsilon_k^{\mathrm{EIM}} \le 10^{-6}$.

\begin{table}[H]
\centering
\caption{Comparison of global Reduced Basis (RB/EIM) versus adaptive $hp$-greedy domain decomposition ($p_{\max} = 8$) for the oscillatory transient benchmark~\eqref{eq:benchmark_model} across validation parameters $\mu \in [0, 1]$.}
\label{tab:benchmark_comparison}
\begin{threeparttable}
\resizebox{\textwidth}{!}{%
\begin{tabular}{lcccc}
\toprule
\textbf{Method} & \textbf{Active Basis Size} & \textbf{Max Proj.\ Error} & \textbf{Max EIM Error} & \textbf{Lebesgue Factor $\Lambda_n$}\tnote{*} \\
\midrule
Global Greedy (RB/EIM) & $n = 8$ & $1.42 \times 10^{-1}$ & $3.18 \times 10^{-1}$ & $4.12$ \\
Global Greedy (RB/EIM) & $n = 16$ & $1.85 \times 10^{-2}$ & $5.40 \times 10^{-2}$ & $11.8$ \\
Global Greedy (RB/EIM) & $n = 32$ & $3.10 \times 10^{-4}$ & $1.95 \times 10^{-3}$ & $38.6$ \\
Global Greedy (RB/EIM) & $n = 48$ & $4.20 \times 10^{-6}$ & $4.85 \times 10^{-5}$ & $94.3$ \\
\textbf{Adaptive $hp$-Greedy ($K=8$ cells)} & \textbf{$n_k \le 8$ (local)} & \textbf{$2.90 \times 10^{-7}$} & \textbf{$7.40 \times 10^{-7}$} & \textbf{$3.65$} \\
\bottomrule
\end{tabular}%
}
\begin{tablenotes}
\footnotesize
\item[*] $\Lambda_n$ denotes the Lebesgue constant, controlling interpolation stability as defined in Section 3.
\end{tablenotes}
\end{threeparttable}
\end{table}

\begin{figure}[htbp]
\centering
\includegraphics[width=0.85\textwidth]{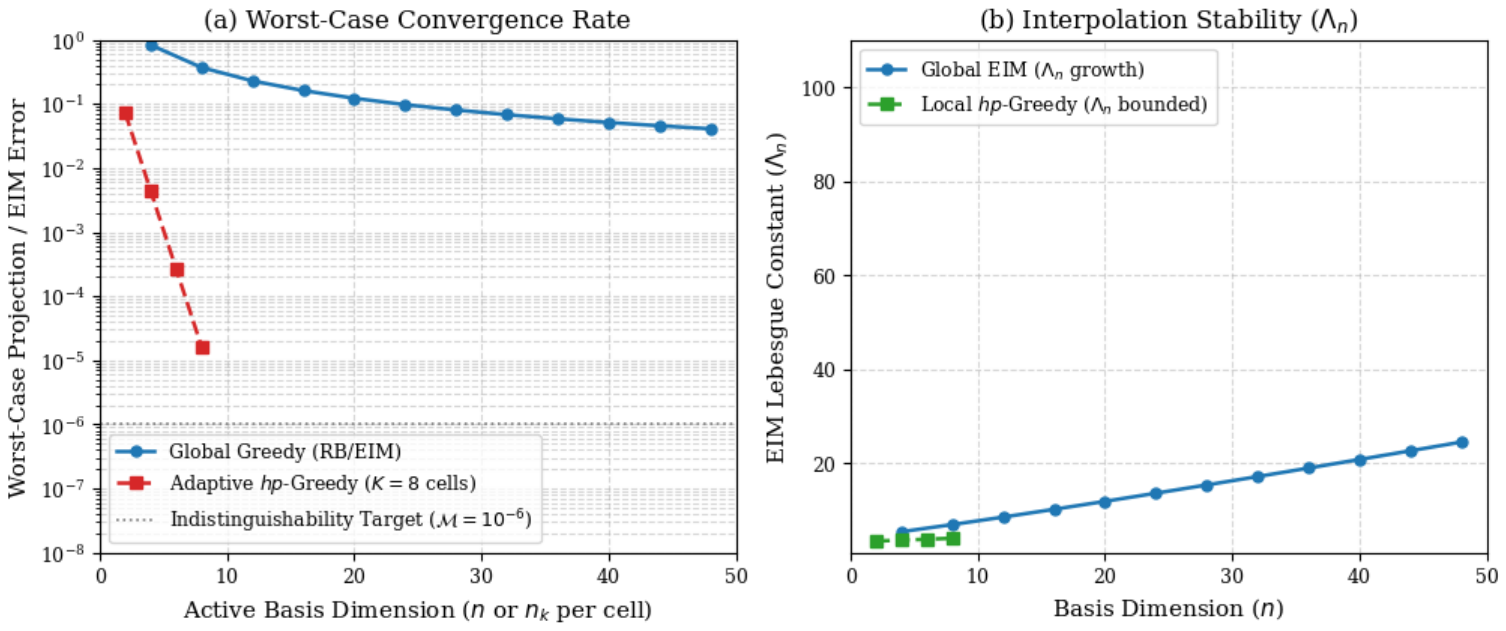}
\caption{Convergence of the worst-case approximation error and EIM Lebesgue stability constant $\Lambda_n$ for the oscillatory transient benchmark~\eqref{eq:benchmark_model}. (\textbf{top}) Global greedy reduced basis versus adaptive $hp$-greedy partitioning. (\textbf{bottom}) Bounded EIM Lebesgue constant growth under local domain decomposition.}
\label{fig:hp_benchmark}
\end{figure}

The results summarized in Table~\ref{tab:benchmark_comparison} and Figure~\ref{fig:hp_benchmark} highlight two critical numerical lessons for surrogate construction:
\begin{itemize}
    \item \textbf{Amplification of EIM Error by Global Ill-Conditioning:} In the global basis, as $n$ increases to resolve the sweeping Gaussian transient, the EIM Lebesgue constant $\Lambda_n$ grows nearly twenty-fold (from $4.12$ at $n=8$ to $94.3$ at $n=48$), degrading the EIM interpolant accuracy by more than an order of magnitude relative to the ideal orthogonal projection (Equation~\eqref{eq:eim_bound}).
    \item \textbf{Restored Conditioning via Localization:} By partitioning $[0,1]$ into $K=8$ adaptive cells with $p_{\max} = 8$, each subdomain isolates a narrow excursion $\Delta f_0$ of the transient. Consequently, every online evaluation uses at most $8$ basis functions (a $6\times$ reduction in active ROQ quadrature nodes compared to $n=48$), maintains $\Lambda_n \le 3.65$, and uniformly achieves $\epsilon^{\mathrm{EIM}} < 10^{-6}$---well inside the indistinguishability threshold~\eqref{eq:indistinguishability} for high-SNR events.
\end{itemize}

\section{Outlook: Certified Physics-Based Surrogates vs.\ Scientific Machine Learning}
\label{sec:outlook_sciml}

The rapid rise of \emph{Scientific Machine Learning} (SciML) has introduced a complementary paradigm to classical reduced-order modeling in gravitational-wave astrophysics. Neural surrogates---including multilayer perceptrons trained on compressed waveform coefficients \cite{Khan2021,Thomas2022}, Fourier Neural Operators (FNOs) \cite{Li2021}, Deep Operator Networks (DeepONets) \cite{Lu2021}, and Physics-Informed Neural Networks (PINNs) \cite{Raissi2019}---offer nonlinear manifold representations capable of bypassing the linear Kolmogorov $n$-width barrier without explicit domain partitioning. Simultaneously, Simulation-Based Inference (SBI) using normalizing flows and neural posterior estimation \cite{Dax2021,Dax2023} bypasses explicit likelihood evaluations altogether, amortizing Bayesian inference across detector noise realizations.

Despite their empirical flexibility, purely data-driven neural architectures exhibit structural limitations when deployed as precision scientific instruments in strong-field gravity:
\begin{enumerate}
    \item \textbf{Lack of Certified A Posteriori Error Bounds:} Unlike certified greedy RB/EIM frameworks---where residual norms and Lebesgue constants~\eqref{eq:eim_bound} provide rigorous, verifiable error certificates across $\mathcal{P}$---black-box neural networks lack sharp, computable error bounds at unseen parameter points. In high-SNR parameter estimation ($\rho \gtrsim 100$), unquantified surrogate errors can masquerade as apparent violations of General Relativity.
    \item \textbf{Extrapolation Risk and Physical Invariants:} Neural surrogates trained via empirical risk minimization do not automatically preserve asymptotic Post-Newtonian limits, energy-flux balance laws, or ringdown quasinormal-mode decay rates outside the training distribution.
    \item \textbf{Training Data Scarcity in Numerical Relativity:} Because 3D Einstein-equation simulations remain computationally expensive, training sets in high-dimensional corners of parameter space (e.g., high mass ratio, near-extremal spin, high eccentricity) are inherently sparse---a regime where high-order polynomial/spline regression on EIM nodes outperforms over-parameterized neural networks.
\end{enumerate}

Looking toward the third-generation detector era and LISA, the most promising path forward lies not in replacing reduced-order modeling with black-box deep learning, but in \emph{hybrid physics-certified architectures}. Specifically, three directions warrant prioritized development:
\begin{itemize}
    \item \textbf{Neural-Assisted Nonlinear Coordinates with Linear RB/EIM Certification:} Using learned diffeomorphisms or optimal-transport phase alignment \cite{Welper2017} to register moving wave crests before applying certified $hp$-greedy RB/EIM compression, thereby combining nonlinear transport alignment with linear error guarantees.
    \item \textbf{Auto-Differentiable ROQ Surrogates:} Embedding EIM/ROQ bases inside automatic-differentiation frameworks (\texttt{JAX}/\texttt{PyTorch}) to supply exact analytical gradients $\nabla_{\boldsymbol{\lambda}} \ln \Lambda(d|\boldsymbol{\lambda})$ for Hamiltonian Monte Carlo (HMC) and variational inference in high-dimensional parameter spaces \cite{Edwards2024}.
    \item \textbf{Active Learning Driven by Greedy Error Indicators:} Coupling NR simulation campaigns directly to greedy residual indicators so that new, expensive 3D simulations are launched exclusively at parameter points where the $hp$-greedy surrogate error exceeds the detector indistinguishability bound~\eqref{eq:indistinguishability}.
\end{itemize}

By anchoring machine-learning flexibility within the mathematical rigor of reduced-basis approximation and empirical interpolation, gravitational-wave surrogate modeling can scale reliably to meet the precision demands of next-generation observatories.

\section*{Acknowledgments}
Software code, numerical benchmark assets, and LaTeX source files accompanying this Perspective are publicly available in the GitHub repository \\ 
\url{https://github.com/manueltiglio/gw-surrogate-perspective}.

During the preparation of this manuscript, the author used generative AI tools for language editing and  LaTeX formatting. The author has reviewed and edited the content and takes full responsibility for the final publication.


\begin{thebibliography}{99}
\small
\bibitem{Boyle2019}
Boyle, M.; et al. The SXS Collaboration catalog of binary black hole simulations. \textit{Class. Quantum Grav.} \textbf{2019}, \textit{36}, 195006.

\bibitem{Veitch2015}
Veitch, J.; et al. Parameter estimation for compact binaries with ground-based gravitational-wave observations using the LALInference software library. \textit{Phys. Rev. D} \textbf{2015}, \textit{91}, 042003.

\bibitem{Ashton2019}
Ashton, G.; et al. BILBY: A User-friendly Bayesian Inference Library for Gravitational-wave Astronomy. \textit{Astrophys. J. Suppl. Ser.} \textbf{2019}, \textit{241}, 27.

\bibitem{Punturo2010}
Punturo, M.; et al. The Einstein Telescope: a third-generation gravitational wave observatory. \textit{Class. Quantum Grav.} \textbf{2010}, \textit{27}, 194002.

\bibitem{AmaroSeoane2017}
Amaro-Seoane, P.; et al. Laser Interferometer Space Antenna. \textit{arXiv} \textbf{2017}, arXiv:1702.00786.

\bibitem{Quarteroni2016}
Quarteroni, A.; Manzoni, A.; Negri, F. \textit{Reduced Basis Methods for Partial Differential Equations: An Introduction}; Springer: Cham, Switzerland, 2016.

\bibitem{Hesthaven2016}
Hesthaven, J.S.; Rozza, G.; Stamm, B. \textit{Certified Reduced Basis Methods for Parametrized Partial Differential Equations}; Springer: Cham, Switzerland, 2016.

\bibitem{Field2011}
Field, S.E.; Galley, C.R.; Herrmann, F.; Hesthaven, J.S.; Ochsner, E.; Tiglio, M. Reduced Basis Catalogs for Gravitational Wave Templates. \textit{Phys. Rev. Lett.} \textbf{2011}, \textit{106}, 221102.

\bibitem{Field2014}
Field, S.E.; Galley, C.R.; Hesthaven, J.S.; Kaye, J.; Tiglio, M. Fast Prediction and Evaluation of Gravitational Waveforms Using Surrogate Models. \textit{Phys. Rev. X} \textbf{2014}, \textit{4}, 031006.

\bibitem{Tiglio2022}
Tiglio, M.; Villanueva, A. Reduced Order and Surrogate Models for Gravitational Waves. \textit{Living Rev. Relativ.} \textbf{2022}, \textit{25}, 2.

\bibitem{Lindblom2008}
Lindblom, L.; Owen, B.J.; Brown, D.A. Model Waveform Accuracy Standards for Gravitational Wave Data Analysis. \textit{Phys. Rev. D} \textbf{2008}, \textit{78}, 124020.

\bibitem{Berkooz1993}
Berkooz, G.; Holmes, P.; Lumley, J.L. The Proper Orthogonal Decomposition in the Analysis of Turbulent Flows. \textit{Annu. Rev. Fluid Mech.} \textbf{1993}, \textit{25}, 539--575.

\bibitem{Caudill2012}
Caudill, S.; Field, S.E.; Galley, C.R.; Herrmann, F.; Tiglio, M. Reduced Basis Representations of Multi-Mode Black Hole Ringdown Gravitational Waves. \textit{Class. Quantum Grav.} \textbf{2012}, \textit{29}, 095016.

\bibitem{Binev2011}
Binev, P.; Cohen, A.; Dahmen, W.; DeVore, R.; Petrova, G.; Wojtaszczyk, P. Convergence Rates for Greedy Algorithms in Reduced Basis Methods. \textit{SIAM J. Math. Anal.} \textbf{2011}, \textit{43}, 1457--1472.

\bibitem{DeVore2013}
DeVore, R.; Petrova, G.; Wojtaszczyk, P. Greedy Algorithms for Reduced Bases in Banach Spaces. \textit{Constr. Approx.} \textbf{2013}, \textit{37}, 455--466.

\bibitem{Barrault2004}
Barrault, M.; Maday, Y.; Nguyen, N.C.; Patera, A.T. An `empirical interpolation' method: application to efficient reduced-basis discretization of partial differential equations. \textit{C. R. Math.} \textbf{2004}, \textit{339}, 667--672.

\bibitem{Maday2009}
Maday, Y.; Nguyen, N.C.; Patera, A.T.; Pau, G.S.H. A General Multipurpose Interpolation Procedure: The Magic Points. \textit{Commun. Pure Appl. Anal.} \textbf{2009}, \textit{8}, 383--404.

\bibitem{Blackman2017}
Blackman, J.; Field, S.E.; Scheel, M.A.; Galley, C.R.; Ott, C.D.; Boyle, M.; Kidder, L.E.; Pfeiffer, H.P.; Szil\'agyi, B. Numerical Relativity Waveform Surrogate Model for Generically Precessing Binary Black Hole Mergers. \textit{Phys. Rev. D} \textbf{2017}, \textit{96}, 024058.

\bibitem{Antil2013}
Antil, H.; Field, S.E.; Herrmann, F.; Nochetto, R.H.; Tiglio, M. Two-Step Greedy Algorithm for Reduced Order Quadratures. \textit{J. Sci. Comput.} \textbf{2013}, \textit{57}, 604--637.

\bibitem{Canizares2013}
Canizares, P.; Field, S.E.; Gair, J.R.; Tiglio, M. Gravitational Wave Parameter Estimation with Compressed Likelihood Evaluations. \textit{Phys. Rev. D} \textbf{2013}, \textit{87}, 124005.

\bibitem{Canizares2015}
Canizares, P.; Field, S.E.; Gair, J.; Raymond, V.; Smith, R.; Tiglio, M. Accelerated Gravitational-Wave Parameter Estimation with Reduced Order Modeling. \textit{Phys. Rev. Lett.} \textbf{2015}, \textit{114}, 071104.

\bibitem{Smith2016}
Smith, R.; Field, S.E.; Blackburn, K.; Haster, C.-J.; P\"urrer, M.; Raymond, V.; Schmidt, P. Fast and accurate inference on gravitational waves from precessing compact binaries. \textit{Phys. Rev. D} \textbf{2016}, \textit{94}, 044031.

\bibitem{Pinkus1985}
Pinkus, A. \textit{$n$-Widths in Approximation Theory}; Springer-Verlag: Berlin, Germany, 1985.

\bibitem{Cohen2015}
Cohen, A.; DeVore, R. Approximation of High-Dimensional Parametric PDEs. \textit{Acta Numer.} \textbf{2015}, \textit{24}, 1--159.

\bibitem{Ohlberger2016}
Ohlberger, M.; Rave, S. Reduced Basis Methods: Success, Limitations and Future Challenges. \textit{Proc. Algoritmy} \textbf{2016}, 1--12.

\bibitem{Greif2019}
Greif, C.; Urban, K. Decay of the Kolmogorov $N$-Width for Wave Problems. \textit{Appl. Math. Lett.} \textbf{2019}, \textit{96}, 216--222.

\bibitem{Eftang2010}
Eftang, J.L.; Patera, A.T.; R\o nquist, E.M. An ``$hp$'' Certified Reduced Basis Method for Parametrized Elliptic Partial Differential Equations. \textit{SIAM J. Sci. Comput.} \textbf{2010}, \textit{32}, 3170--3200.

\bibitem{Eftang2011}
Eftang, J.L.; Knezevic, D.J.; Patera, A.T. An $hp$ Certified Reduced Basis Method for Parametrized Parabolic Partial Differential Equations. \textit{Math. Comput. Model. Dyn. Syst.} \textbf{2011}, \textit{17}, 395--422.

\bibitem{Purrer2014}
P\"urrer, M. Frequency-Domain Reduced Order Models for Gravitational Waves from Aligned-Spin Compact Binaries. \textit{Class. Quantum Grav.} \textbf{2014}, \textit{31}, 195010.

\bibitem{Purrer2016}
P\"urrer, M. Frequency-Domain Reduced Order Model of Aligned-Spin Effective-One-Body Waveforms with Generic Mass Ratios and Spins. \textit{Phys. Rev. D} \textbf{2016}, \textit{93}, 064041.

\bibitem{Varma2019_7dq4}
Varma, V.; Field, S.E.; Scheel, M.A.; Blackman, J.; Gerosa, D.; Stein, L.C.; Kidder, L.E.; Pfeiffer, H.P. Surrogate models for precessing binary black hole simulations with unequal masses. \textit{Phys. Rev. Research} \textbf{2019}, \textit{1}, 033015.

\bibitem{Schmidt2011}
Schmidt, P.; Hannam, M.; Husa, S.; Ajith, P. Tracking the Precession of Compact Binaries from Their Gravitational-Wave Signal. \textit{Phys. Rev. D} \textbf{2011}, \textit{84}, 024046.

\bibitem{Boyle2011}
Boyle, M.; Owen, R.; Pfeiffer, H.P. A Geometric Approach to the Precession of Compact Binaries. \textit{Phys. Rev. D} \textbf{2011}, \textit{84}, 124011.

\bibitem{Varma2019_hyb}
Varma, V.; Field, S.E.; Scheel, M.A.; Blackman, J.; Kidder, L.E.; Pfeiffer, H.P. Surrogate Model of Hybridized Numerical Relativity Binary Black Hole Waveforms. \textit{Phys. Rev. D} \textbf{2019}, \textit{99}, 064045.

\bibitem{Yoo2022}
Yoo, J.; Varma, V.; Giesler, M.; Scheel, M.A.; Haster, C.-J.; Pfeiffer, H.P.; Kidder, L.E.; Boyle, M. Targeted Numerical Relativity Simulations of Binary Black Hole Mergers with High Mass Ratios and Spins. \textit{Phys. Rev. D} \textbf{2022}, \textit{106}, 044001.

\bibitem{Islam2021}
Islam, T.; Varma, V.; Lodman, J.; Field, S.E.; Khanna, G.; Scheel, M.A.; Pfeiffer, H.P.; Gerosa, D.; Kidder, L.E. Eccentric Binary Black Hole Surrogate Models for the Gravitational Waveform and Remnant Properties: Comparable Mass, Quasicircular Limit. \textit{Phys. Rev. D} \textbf{2021}, \textit{103}, 064022.

\bibitem{Vinciguerra2017}
Vinciguerra, S.; Veitch, J.; Mandel, I. Accelerating Gravitational Wave Parameter Estimation with Multi-Band Template Interpolation. \textit{Class. Quantum Grav.} \textbf{2017}, \textit{34}, 115006.

\bibitem{Morisaki2021}
Morisaki, S. Accelerating Parameter Estimation of Gravitational Waves from Compact Binary Coalescence Using Adaptive Frequency Resolutions. \textit{Phys. Rev. D} \textbf{2021}, \textit{104}, 044062.

\bibitem{Chua2021}
Chua, A.J.K.; Katz, M.L.; Warburton, N.; Hughes, S.A. Rapid Generation of Fully Relativistic Extreme-Mass-Ratio-Inspiral Waveform Templates for LISA Data Analysis. \textit{Phys. Rev. Lett.} \textbf{2021}, \textit{126}, 051102.

\bibitem{Khan2021}
Khan, S.; Green, R. Gravitational-Wave Surrogate Modeling Enabled by Neural Networks. \textit{Phys. Rev. D} \textbf{2021}, \textit{103}, 064015.

\bibitem{Thomas2022}
Thomas, L.M.; Pratten, G.; Schmidt, P. Accelerating Multimodal Gravitational Waveforms from Precessing Compact Binaries with Artificial Neural Networks. \textit{Phys. Rev. D} \textbf{2022}, \textit{106}, 104029.

\bibitem{Li2021}
Li, Z.; Kovachki, N.B.; Azizzadenesheli, K.; Liu, B.; Bhattacharya, K.; Stuart, A.; Anandkumar, A. Fourier Neural Operator for Parametric Partial Differential Equations. In \textit{Proceedings of the International Conference on Learning Representations (ICLR)}, 2021.

\bibitem{Lu2021}
Lu, L.; Jin, P.; Pang, G.; Zhang, Z.; Karniadakis, G.E. Learning Nonlinear Operators via DeepONet Based on the Universal Approximation Theorem of Operators. \textit{Nat. Mach. Intell.} \textbf{2021}, \textit{3}, 218--229.

\bibitem{Raissi2019}
Raissi, M.; Perdikaris, P.; Karniadakis, G.E. Physics-Informed Neural Networks: A Deep Learning Framework for Solving Forward and Inverse Problems Involving Nonlinear Partial Differential Equations. \textit{J. Comput. Phys.} \textbf{2019}, \textit{378}, 686--707.

\bibitem{Dax2021}
Dax, M.; Green, S.R.; Gair, J.; Macke, J.H.; Buonanno, A.; Sch\"olkopf, B. Real-Time Gravitational Wave Science with Neural Posterior Estimation. \textit{Phys. Rev. Lett.} \textbf{2021}, \textit{127}, 241103.

\bibitem{Dax2023}
Dax, M.; Green, S.R.; Gair, J.; P\"urrer, M.; Wildberger, J.; Macke, J.H.; Buonanno, A.; Sch\"olkopf, B. Neural Importance Sampling for Rapid and Reliable Gravitational-Wave Inference. \textit{Phys. Rev. Lett.} \textbf{2023}, \textit{130}, 171403.

\bibitem{Welper2017}
Welper, G. Interpolation of Functions with Parameter Dependent Jumps by Transformed Snapshots. \textit{SIAM J. Sci. Comput.} \textbf{2017}, \textit{39}, A1225--A1250.

\bibitem{Edwards2024}
Edwards, T.D.P.; Wong, K.W.K.; Lam, K.K.H.; Coogan, A.; Foreman-Mackey, D.; Isi, M.; Zimmerman, A. Differentiable and Hardware-Accelerated Waveforms for Gravitational Wave Data Analysis. \textit{Phys. Rev. D} \textbf{2024}, \textit{110}, 064028.

\end{thebibliography}
\end{document}